\documentclass[12pt,preprint]{aastex}
\slugcomment{Accepted for publication in ApJ.}

\usepackage{natbib}
\usepackage{textcomp}

\shorttitle{Main Belt Comet 313P/Gibbs}
\shortauthors{Pozuelos et al.}

\begin{document}


\title{On the dust environment of Main-Belt Comet 313P/Gibbs}


\author{F.J. Pozuelos\altaffilmark{1}, A. Cabrera-Lavers\altaffilmark{2,3}, J. Licandro\altaffilmark{2,3} and F. Moreno\altaffilmark{1}}
\affil{$^{1}$ Instituto de Astrof\'isica de Andaluc\'ia, CSIC,
    Glorieta de la Astronom\'ia s/n, 18008 Granada, Spain.\\
    $^{2}$ Instituto de Astrof\'isica de Canarias, c/V\'ia L\'actea s/n, E-38200 La Laguna, Tenerife, Spain.\\
    $^{3}$ Departamento de Astrof\'isica, Universidad de la Laguna (ULL), E-38205 La Laguna, Tenerife, Spain.\\}
\email{pozuelos@iaa.es}

%
%




\begin{abstract}
We present observations carried out using the 10.4 m Gran Telescopio Canarias 
and an interpretative model of the dust environment of activated asteroid 313P/Gibbs.
We discuss three different models relating to different values of the dust parameters, i.e, dust 
loss rate, maximum and minimum sizes of particles, power index of the size distribution, and 
emission pattern. The best model corresponds with an isotropic emission of particles which started
on August 1st. The size of grains were in the range of $0.1-2000$ $\mu$m, with velocities
for 100 $\mu$m particles between $0.4-1.9$ m$~$s$^{-1}$, with a dust
production rate in the range of $0.2-0.8$ kg$~$s$^{-1}$.
The dust tails' brightness and morphology are best interpreted in terms of a model of sustained
and low dust emission driven by water-ice sublimation, spanning since 2014 August 1st, and triggered by
a short impulsive event. This event produced an emission of small particles of about 0.1 $\mu$m with 
velocities of $\sim$4 m$~$s$^{-1}$. From our model we deduce that the activity of this Main-Belt Comet continued for, 
at least, four months, since activation.

\end{abstract}


\keywords{comets: individual (313P/Gibbs) --- methods: numerical --- minor planets, asteroids: general}



\section{INTRODUCTION}

The Main-Belt Comet (MBC) 313P/Gibbs, hereafter 313P, was discovered
in the course of the Catalina Sky Survey's 0.68-m Schmidt telescope on September 24.3, 
2014. This MBC was described as a faint object with a narrow tail 20\arcsec~ in 
length and position angle 270$\degr$ \citep{gibbs2014}.  
These objects have the dynamical 
properties of asteroids (i.e., Tisserand parameter with respect to Jupiter larger than 3), 
but display comet-like structure due to dust and/or gas emission.

The origin of the activity in these objects is still unclear. A variety of mechanisms have been 
proposed. While some of them are associated with impulsive events, such as collisions and rotational 
break-up \citep[see,e.g.,][]{moreno2014,jewitt2014a,yang2011,ishiguro2011}
others show an activity sustained over time probably due 
to water-ice sublimation \citep[][]{jewitt2014b,moreno2013,licandro2013a,hsieh2012c,moreno2011}.
Of these driving mechanisms of the MBCs activity, it seems that the most likely is water-ice sublimation,
and an attempt to link this fact with the Earth's oceans has been established. Due to the high 
temperatures on Earth when it was formed, an external source of 
water is widely accepted. Comets were considered the best candidates for this theory, but 
the measurements of the isotope ratios of deuterium to hydrogen (D/H ratio) have shown
a wide range of values, but most of them being higher than the oceanic D/H ratio \citep{altwegg2015}. The existence 
of water in the Main-Belt should not be a surprise. Actually, half of the outer belt asteroids show absortion features,
which can be attributed to the presence of hydrated minerals, and water ice has been detected in (24) Thermis.

N-Body integrations have shown that MBCs are dynamically stable on time scales of 100 Myr or longer
\citep[][]{ipatov1999,hsieh2012a,hsieh2012b,stevenson2012}, and, consequently, 
they are considered native members of the Main Asteroid Belt and not 
captured objects from elsewhere \citep{hsieh2009}.
For further details on those objects we 
refer the readers to the works of \citet{bertini2011} and \citet{jewitt2012}. 

Observations and analysis of 313P have recently been carried out by \citet{jewitt2015}, who
report an equilibrium sublimation of dirty water-ice from a small patch of the surface as 
origin of the dust tail. The authors ruled out other possible explanations for the origin
of the activity. Also, \citet{hsieh2015} have studied this object, reaching the same conclusion
that the activity is likely sublimation-driven. In addition, \citet{hui2015} presented 
prediscovery observations of 313P in 2003 and 2004, establishing that, after 133P/Elst-Pizarro and 238P/Read, this MBC is 
the third activated asteroid to show mass loss in different orbits.
In this work, we report images taken at 10.4 m Gran Telescopio Canarias
(GTC) of 313P on three different dates. We use our Monte Carlo dust tail code, in order to 
provide estimates of the ejected mass, the particle ejection velocities, and their size 
distribution. Finally, we compare our results to those of \citet{jewitt2015} and \citet{hsieh2015}.

\section{OBSERVATIONS AND DATA REDUCTION}  

\begin{figure}
\epsscale{.80}
\plotone{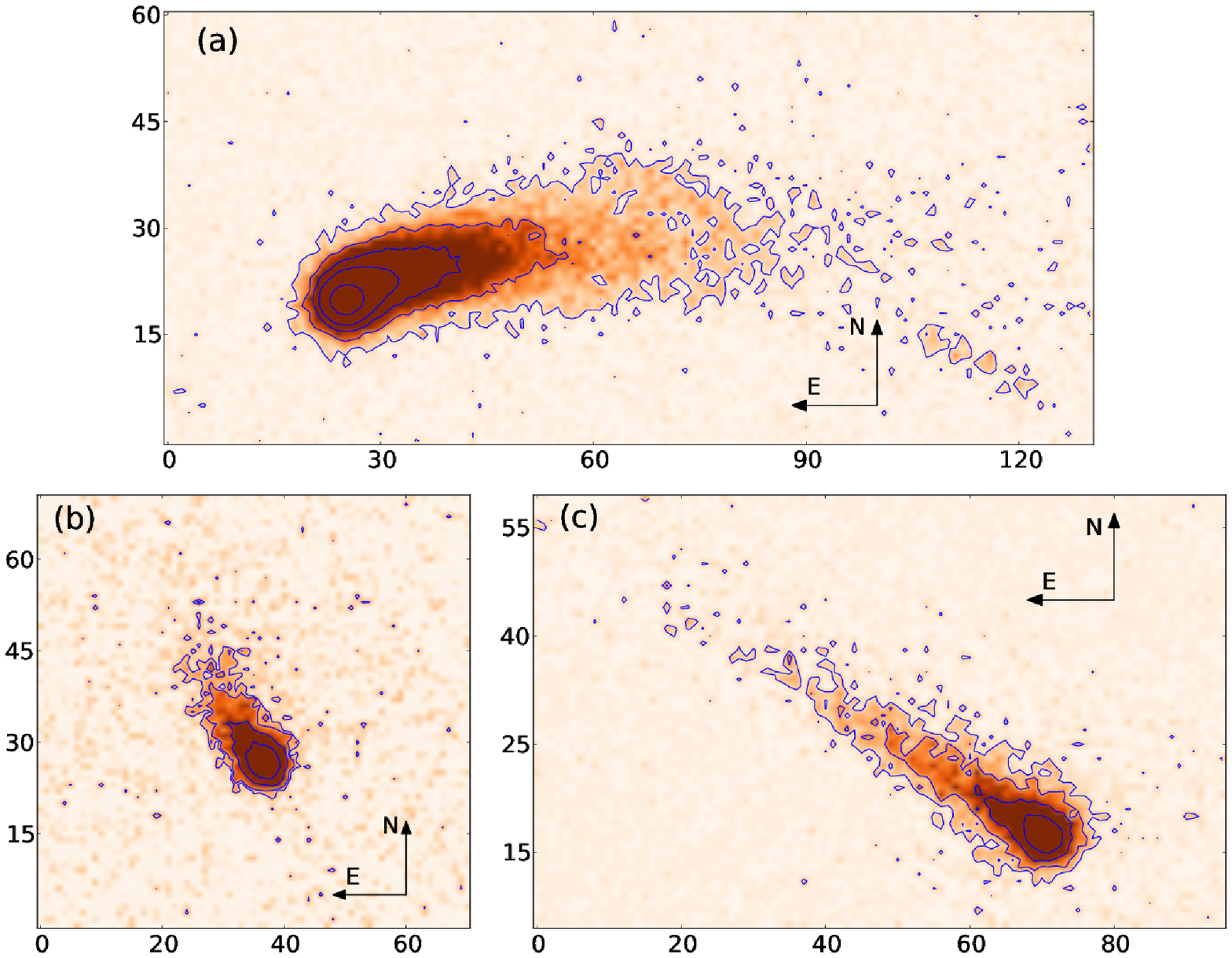}
\caption{Panel (a) displays the 313P/Gibbs imaged on 2014 September 29.07. Isophote levels (solid blue lines)
are 0.25$\times10^{-14}$, 0.65$\times10^{-14}$, 1.50$\times10^{-14}$, 0.35$\times10^{-13}$, and 1.50$\times10^{-13}$ (SDU). The scale is 263 km pixel$^{-1}$. Panel 
(b) shows the same as (a) but for the 2014 November 4.05. Isophote levels are 
1.00$\times10^{-14}$, 1.50$\times10^{-14}$, 3.00$\times10^{-14}$, and 6.00$\times10^{-14}$ (SDU). The scale is 288 km pixel$^{-1}$. Panel (c) 
corresponds with 2014 December 16.85. Isophote levels are 
0.28$\times10^{-14}$, 0.5$\times10^{-14}$, 1.00$\times10^{-14}$, and 1.75$\times10^{-14}$ (SDU). The scale is 373 km pixel$^{-1}$.
In all images North is up and East is to the left.
 \label{f1}}
\end{figure}

The observations of 313P were carried out on the nights of September 29, 
November 4 and December 16 2014 (hereafter 29Sept, 4Nov and 16Dec), using Sloan \emph{r\arcmin} and \emph{g\arcmin} 
filters in the Optical System for Image and Low Resolution Integrated Spectroscopy
(OSIRIS) camera-spectrograph \citep[][]{cepa2000,cepa2010} at the GTC. OSIRIS 
provides a field of view 7\arcmin.8~$\times$~7\arcmin.8 and a pixel scale of
0\arcsec.125 pixel$^{-1}$. In order to increase the signal to noise ratio the 
data were binned in 2~$\times$~2 pixels, so that the spatial resolution of the 
images become 263 km pixel$^{-1}$, 288 km pixel$^{-1}$, and 373 km pixel$^{-1}$
on the observation dates. The images were bias and flat corrected using standard 
techniques and the calibration was performed using
photometric zero points determined by standard star observations. The \emph{r\arcmin}
images of the 313P were transformed into \emph{R} standard Johnson-Cousins magnitude, 
where we assume the same spectral dependence for both the object and the Sun within the 
bandpasses of these two red filters. Then, images were converted to solar disk 
intensity units (SDU) appropriate for analysis in terms of the dust tail models. 
The resulting images and the observational details are shown in Fig. \ref{f1} and Table \ref{T1}, respectively.




\section{THE MODEL}

The analysis was performed using our Monte Carlo dust
tail code, which allows us to generate synthetic dust tail brightness images
to be compared with the observations. This code
has been successfully used in previous works on characterization
of dust environments of MBCs \citep[see, e.g.,][]{moreno2014,moreno2013,moreno2011}
and Jupiter Family Comets \citep[][]{moreno2012,pozuelos2014a,pozuelos2014b}.
In that code, we computed the motion of the particles after they had been ejected from the
nucleus and submitted to the gravity force of the Sun and the radiation pressure,
describing a Keplerian orbit around the Sun. The type of orbit of the dust
particle (elliptical, hyperbolic, or parabolic) is a function of
its ejection velocity and the $\beta$ parameter \citep[e.g.,][]{fulle1989}. 
This parameter
can be expressed as $\beta=C_{pr}Q_{pr}/(2\rho d)$, where
$C_{pr}=1.19\times10^{-3}$ kg m$^{-2}$, $Q_{pr}$ is the radiation pressure
coefficient, $\rho$ is the particle density, and $d$ is the particle diameter.
Owing to the many model inputs, we set some of them to a specific value. Thus, 
particles are considered spherical, with density assumed at $\rho=1000$ kg m$^{-3}$,
and their refractive index is set at 1.88~$+$~0.71$i$, which corresponds with 
carbonaceous composition \citep{edoh1983}. Using the Mie theory for spherical particles, 
we find that geometric albedo is $p_{v}=0.04$, and that $Q_{pr}\sim1$ for particles
of radius $r\geq1$~$\mu$m (\citet{moreno2012b}, their Figure 5). 

The particle size
distribution is assumed to follow a time-dependent power law with index $\alpha$.
This index is within the range of -4.2 to -3, found in a number of comets 
(see, e.g., \citet{jockers1997} and the references therein)
The particle 
ejection velocities are described as a function of $\beta$ parameter as 
$v(\beta)=v_{0}\beta^{\gamma}$. This relationship is generally accepted for 
terminal velocities of comet dust, and also for fragments ejected after collision experiments
\citep[see, e.g.,][]{giblin1998,onose2004}. Thus, $\gamma$, and the time-dependent
parameters, i.e., $v_{0}$, the dust mass loss rate,
the size of particles ($r_{max}$ and $r_{min}$), and the index $\alpha$ of the particle size distribution
as a function of the heliocentric distance are fitting parameters.







\section{RESULTS}

In order to fit the available observations of 313P,
all the parameters described above must be introduced as inputs of the code. 
In order to explore the largest number of possible scenarios, we set $\gamma$ to three different values, i.e., 
 $\gamma=1/2,1/8,$ and $1/20$, and looked for the best fit model in each case, varying the rest of the parameters. 
The general procedure consists of 
a trial-and-error procedure, starting 
from the most simple scenario, where we consider an isotropic 
ejection outgassing model, having minimum and maximum particle
radius of $r_{min}=1$ $\mu$m, $r_{max}=1$ cm, and set $\alpha=-3.5$.
Regarding the dust mass loss rate and $v_{0}$, we impose a monotonically symmetric 
evolution with respect to perihelion. Then, we start to vary the parameters until an 
acceptable agreement with the dust tail images is reached. 

Our first model corresponds with $\gamma=1/2$ (model I). This is consistent with previous studies of the MBCs P/2010 R2 (La Sagra)
\citep{moreno2011} and P/2012 T1 (PANSTARRS) \citep{moreno2013}. After a number of attempts, the best fit model
consists of an emission of particles with sizes ranging between 1-2000 $\mu$m. The power index of the size 
distribution is set to a constant value of $\alpha=-3.2$. The terminal velocities for particles of 100 $\mu$m are between
0.6-5 m$~$s$^{-1}$. The maximum dust production rate is 1 kg$~$s$^{-1}$ and the total dust
emitted is estimated to be 3.2$\times10^{6}$ kg. The emission of particles started at $r_{h}=-2.396$ AU, i.e., August 1st, and
the maximum values for the dust production rate, terminal velocities and the $r_{max}$ are reached 
at $r_{h}=+2.395$ AU, i.e., September 21st.

The second proposed model corresponds to $\gamma=1/8$ (model II). This assumption was also used in \citet{moreno2014} where the
authors studied the activated asteroid P/2013 P5 (PANSTARRS). The range of size particles is 0.1-2000 $\mu$m, with the
values of the power index of the size distribution between $\alpha=-3.1$-$(-3.3)$. The terminal velocities are in the 
range of 0.4-1.9 m$~$s$^{-1}$ for particles of 100 $\mu$m. The peak of the dust production rate is 0.8 kg$~$s$^{-1}$ and the total dust ejected is
3.4$\times10^{6}$ kg. In this model, the emission started at the same time as the model of $\gamma=1/2$, i.e, 
at $r_{h}=-2.396$ AU (August 1st) and the maximum value of the dust parameters were reached also at 
$r_{h}=+2.395$ AU (September 21st). Despite starting and reaching the maximum values for the dust parameters at the same time 
as the model I (August 1st and September 21st respectively), the evolution of those parameters as a function of the heliocentric distance
were different. The terminal velocities for particles
of 100 $\mu$m at the beginning 
of the emission reached the same values as those reached at $r_{h}=+2.395$ AU (September 21st). In addition, the smallest
grains (0.1 $\mu$m) are ejected just at the start of the activity. This points to an emission of particles at high velocities at the beginning 
of the activity.

The last model in our study set $\gamma=1/20$ (model III). This option has been used before in the study of the MBC P/2012 F5 (Gibbs) 
\citep{moreno2012b} and (596) Sheila \citep{moreno2011b}. The size of particles were in the range of 0.1-2000 $\mu$m and 
$\alpha$ took values between $-3.1$-$(-3.2)$. In this case, the terminal velocities of 100 $\mu$m particles were slower than 
in the other two models, with values of 0.4-0.9 m$~$s$^{-1}$. The maximum dust production rate reached the value of 
0.80 kg$~$s$^{-1}$, being the total emitted dust 2.9$\times10^{6}$ kg. As before, the start of the activity seems to be at 
$r_{h}=-2.396$ AU (August 1st), but in this case, the maximum values of the dust parameter were reached at perihelion, i.e.,
August 28th ($r_{h}=\pm 2.3916$ AU).

\begin{figure}
\includegraphics[scale=.62]{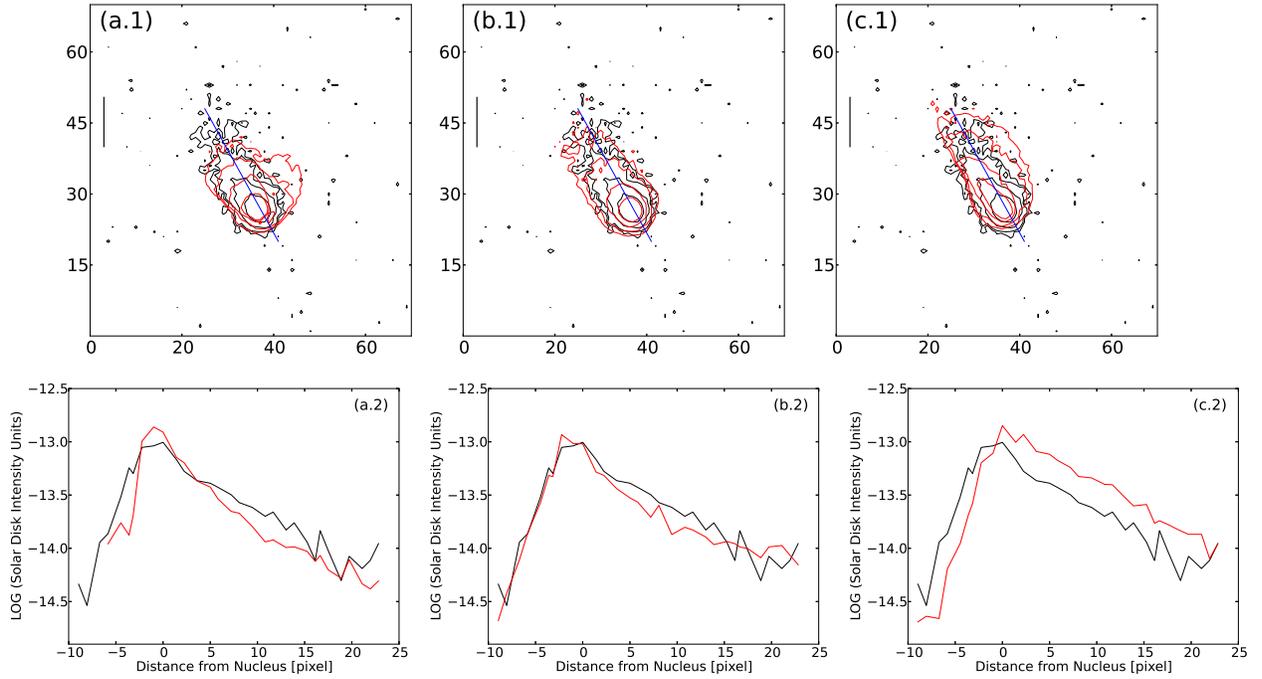}
\caption{Comparison between models on the observation date of 2014 November 4.05. Panels (a.1) and (a.2) correspond to the model with $\gamma=1/2$, 
(b.1) and (b.2) with $\gamma=1/8$, and (c.1) and (c.2) with $\gamma=1/20$. In all cases, the black line 
is the observed image and the red line the fit. The blue line along the tails in panels (a.1), (b.1) and (c.1) correspond
to the brightness scan displayed in (a.2), (b.2) and (c.2). The isophote levels are 
1.00$\times10^{-14}$, 1.50$\times10^{-14}$, 3.00$\times10^{-14}$, and 6.00$\times10^{-14}$ (SDU).
The vertical bars correspond to 3000 km in the sky.
 \label{f2}}
\end{figure}

\begin{figure}
\includegraphics[scale=1.3]{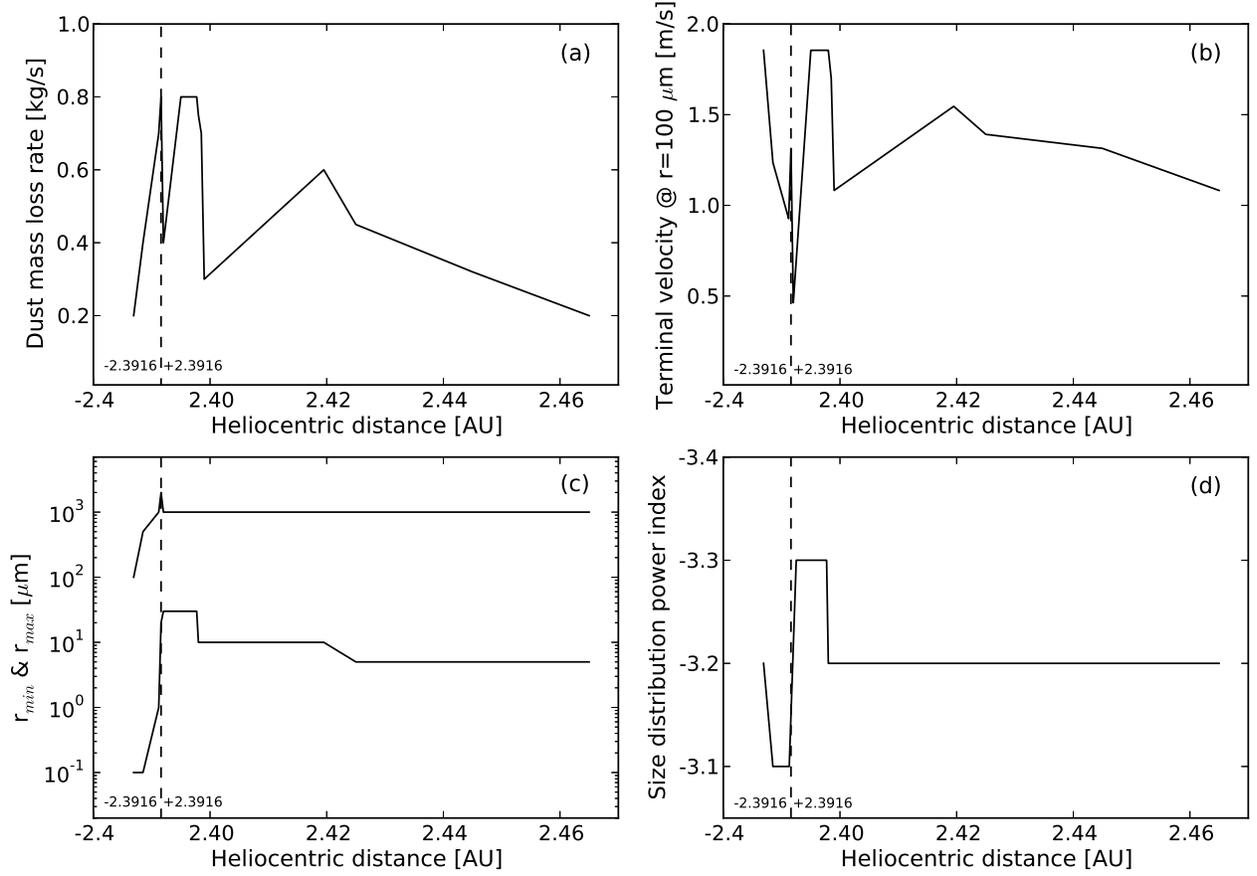}
\caption{Evolution of the dust parameters
obtained in the model II ($\gamma=1/8$) versus the heliocentric distance for
MBC 313P/Gibbs. In panel a) dust mass loss rate [kg$~$s$^{-1}$]; b) terminal velocities for 100 $\mu$m glassy carbon spheres [m$~$s$^{-1}$]; c) maximum and minimum size
of ejected particles [$\mu$m]; d) power index of the size distribution $\alpha$. The dashed vertical line corresponds to
perihelion.\label{fig3}}
\end{figure}

\begin{figure}
\includegraphics[scale=.63]{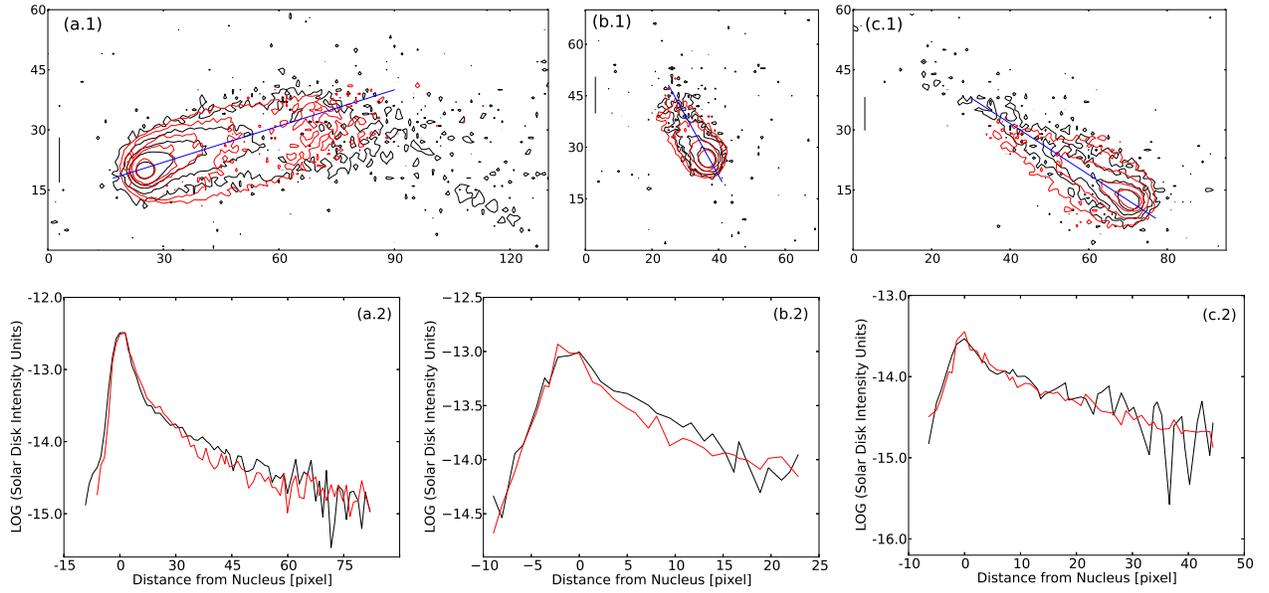}
\caption{Isophote field and brightness scans comparison between observations
and model II. The black lines correspond
to the observations and the red ones to
the model. The blue lines along the tails correspond to the
brightness scans. Panel (a.1) and (a.2) refer to image on September 29.07 2014. The isophote levels are
0.25$\times10^{-14}$, 0.65$\times10^{-14}$, 1.50$\times10^{-14}$, 0.35$\times10^{-13}$, and 1.50$\times10^{-13}$ (SDU). 
Panel (b.1) and (b.2) correspond to the observation on November 4.05 2014. Isophote levels are 
1.00$\times10^{-14}$, 1.50$\times10^{-14}$, 3.00$\times10^{-14}$, and 6.00$\times10^{-14}$ (SDU). Panel (c.1) and (c.2) refer to the observation on 2014 December 16.85. 
Isophote levels are 0.28$\times10^{-14}$, 0.5$\times10^{-14}$, 1.00$\times10^{-14}$, and 1.75$\times10^{-14}$ (SDU).
 \label{fig4}}
\end{figure}

In order to determine which of the proposed models give the best fit, we computed the fitting quality to the observations
by the quantity $\sigma=\sqrt{\sum(I_{obs}-I_{mod})^{2}/N}$, where $N$ is the number of pixels of the images, $I_{obs}$ and $I_{mod}$ are the 
observed and fitted images. The $\sigma$ parameters at each model and date are given in Table \ref{T2}. In addition, in order to 
facilitate an easy inspection of the models, we display in Figure \ref{f2} the comparison between models and observation on November 4.05.
We see that model II gives the lowest value of $\sigma$ at all three dates of observation, and, therefore, this model is favored against models I or III, although
none of those models can be excluded. In Figure \ref{fig3} 
we display the evolution of the dust parameters as a function of heliocentric distance, and 
in Figure \ref{fig4} we present the comparison of this model with the observational data.


\section{DISCUSSION}

From our analysis of the dust environment of 313P, we have shown that the model having $\gamma=1/8$ is favored respect to $\gamma=1/2$ or
$\gamma=1/20$ models. This indicates a very weak dependence 
of $v(\beta)$ with $\beta$. The size of particles is
in the range of 0.1-2000 $\mu$m. The smallest grains in that range were ejected just at the beginning of the 
emission at $r_{h}=-2.396$ AU (August 1st) with terminal velocities of 4.4 m$~$s$^{-1}$, i.e., ejection of small particles
at moderately high velocities at the beginning of the activity. In general terms, the peak of the activity is reached 
at $r_{h}=+2.395$ AU (September 21st). The dust production rate was in the range of 0.2-0.8 kg$~$s$^{-1}$, which is
weak activity, but it lasted about four months from the start of the emission until our last observation on December 16.85, 
with a total of dust emitted of $3.4\times10^{6}$ kg.
This emission is consistent with the sublimation of volatile ices found in previous works, such as
P/2005 U1 (Read) in 2005, when the comet displayed a substantial coma with large particles of $r\gtrsim100$ $\mu$m,
and ejection velocities of $\sim0.2-3$ m s$^{-1}$ \citep{hsieh2009}.
Therefore, we conclude that 313P activity is sustained probably by water-ice sublimation triggered by a short impulsive event, 
like a collisional excavation of surface ice.

\citet{jewitt2015}, obtain a continuous mass-loss 
at rates of $\sim$0.2-0.4 kg s$^{-1}$, with size particles of 50-100~$\mu$m.
The maximum values of those ranges are a bit lower than ours, but close to them. 
The authors concluded that the ejection of material is consistent 
with the action of sublimation of dirty water-ice from an exposed 
patch covering only a few hundred square meters, and protracted from September until November.  In our model we do not 
constrain the activity to an active area, but the emission of particles can be explained by an isotropic ejection model. 
In addition, \citet{hsieh2015} gave a 
total of dust ejected in the range of 2-4$\times10^{7}$ kg, which is larger than the dust production obtained in our model.
Regarding the origin of the activity of the MBC, both \citet{jewitt2015} and \citet{hsieh2015}, 
suggest an impact or local avalanche as the nature of the trigger to expose ice in first place, which is consistent 
with the results obtained in our model. 

In addition, \citet{hui2015} report the existence of archival data showing that 313P was active in its perihelion 
passage in 2003, favoring the theory of sublimation of water-ice. This puts 313P
in the same category of active asteroids as 133P/Elst-Pizarro and 238P/Read, 
which exhibit activity at more than one perihelion passage \citep[][]{hsieh2004,hsieh2011,hsieh2013,moreno2013}.
This fact is considered the most reliable indicator of water-ice sublimation.
Despite this evidence, until today, no gas has been spectroscopically detected in any MBC, 
making the theory of water-ice sublimation incomplete \citep[see, e.g.,][]{valborro2012,licandro2011b,jewitt2009}.
Therefore, the driver of the activity in 313P can not be clearly established.

\section{CONCLUSIONS}

The application of the Monte Carlo dust tail model to images of the 
recently discovered MBC 313P/Gibbs, carried out at GTC, has revealed that the observed
morphology and evolution of the tail is consistent with the emission of particles started 
on August 1st ($r_{h}=-2.396$ AU), 
with particle sizes in the range of 0.1-2000 $\mu$m, which were ejected isotropically. 
The dust loss rate varies between 0.2-0.8 kg$~$s$^{-1}$, reaching the 
peak of the activity at $r_{h}=+2.395$ AU (September 21st). The terminal velocities are parametrized by 
$v(\beta)=v_{0}\beta^{\gamma}$, where we favor $\gamma=1/8$ over the other tested values ($\gamma=1/2$ and $\gamma=1/20$). 
Those velocities showed that at the beginning of the emission, the smallest grains (0.1 $\mu$m) 
were moderately rapidly dispersed, with velocities of 4.4 m$~$s$^{-1}$. The total dust ejected is 
found to be 3.4$\times10^{6}$ kg. Therefore, we conclude that
the sustained low activity observed for a few months is likely driven by water-ice sublimation, due to a collisional event,
which exposed fresh ices to the solar radiation. This scenario and our estimates of the dust production rates, 
ejection velocities, and particle sizes are in agreement with the results
recently reported by \citet{jewitt2015} and \citet{hsieh2015} for this MBC.

\acknowledgments
We thank an anonymous referee for comments and suggestions for improving the paper.

This article is based on observations made with the Gran Telescopio Canarias (GTC), 
instaled in the Spanish Observatorio del Roque de los Muchachos of the Instituto de Astrof\'isica de Canarias, 
in the island of La Palma.
 
This work was supported by contracts AYA2012-
3961-CO2-01 and FQM-4555 (Proyecto de Excelencia, Junta de Andalucia).

\clearpage
\bibliographystyle{apj}
\bibliography{curroreferencias}

\clearpage



\clearpage



\clearpage

\begin{deluxetable}{ccccccc}
\tablecaption{Log of the Observations.\label{T1}}
 \tablewidth{0pt}
 \tablehead{
 \colhead{Observation Date} & \colhead{ Days from} & \colhead{$r_{h}$} & \colhead{$\Delta$} & \colhead{Phase} & \colhead{Position} & \colhead{Resolution} \\
 \colhead{(UT) } & \colhead{perihelion\tablenotemark{a} } & \colhead{(AU)} & \colhead{(AU)} & \colhead{angle~($\degr$)} & \colhead{angle~($\degr$)} & \colhead{(km pixel$^{-1}$)}
 }
 \startdata
2014 Sept. 29.07 & 31.6  & 2.398 & 1.429  & 7.9   & 317.7 & 263  \\
2014 Nov. 4.05   & 67.5  & 2.419 & 1.565  & 14.8  & 43.5  & 288  \\
2014 Dec. 16.85  & 110.3 & 2.464 & 2.030  & 22.7  & 63.2  & 373  \\
 \enddata
\tablenotetext{a}{All these observations are post-perihelion.}
\end{deluxetable}

\begin{deluxetable}{ccccc}
\tablecaption{Fitting quality of the observed images.\label{T2}}
 \tablewidth{0pt}
 \tablehead{
 \colhead{ Model } & \colhead{$\gamma$} & \colhead{2014 Sept. 29.07} & \colhead{2014 Nov. 4.05} & \colhead{2014 Dec. 16.85} \\
 }
 \startdata
I & 1/2  & 1.2$\times10^{-14}$ & 9.7$\times10^{-15}$  & 3.3$\times10^{-15}$  \\
II   & 1/8  & 8.2$\times10^{-15}$ & 8.1$\times10^{-15}$  & 2.3$\times10^{-15}$  \\
III  & 1/20 & 2.2$\times10^{-14}$ & 1.2$\times10^{-14}$  & 2.6$\times10^{-15}$  \\
 \enddata
\end{deluxetable}

\clearpage

\end{document}